\documentclass[10pt,conference,compsocconf]{IEEEtran}
\usepackage{times}

\usepackage{caption}
\usepackage{listings}
\usepackage{amsmath}
\usepackage{drawstack}
\usepackage{tikz}
\usetikzlibrary{er,positioning, shapes, shapes.geometric, arrows, trees, shapes.multipart, calc}
\tikzstyle{process} = [rectangle, text width=10em,minimum height = .7cm,text centered, draw=black, fill=white]
\tikzstyle{data} = [rectangle,rounded corners,text width=10em,minimum height = .7cm,text centered, draw=black, fill=gray]

\newcommand{\apate}{Apate }
\newcommand{\apaten}{Apate}
\newcommand{\hih}{High Interaction Honeypot }
\newcommand{\hihn}{High Interaction Honeypot}

\newcommand{\ids}{Intrusion Detection }

\newcommand{\itsec}{IT Security }

\newcommand{\syscall}[1]{{\em sys\_#1 }}
\newcommand{\syscalln}[1]{{\em sys\_#1}}
\newcommand{\ita}[1]{{\em #1 }}
\newcommand{\itan}[1]{{\em #1}}

\newenvironment{formelnon}{
\begin{equation}
\begin{aligned}
}
{\end{aligned}\end{equation}}

\begin{document}
\pagenumbering{gobble}
%

\title{\textbf{\Large Apate - A Linux Kernel Module for High Interaction Honeypots}\\[0.2ex]}

\author{\IEEEauthorblockN{~\\[-0.4ex]\large Christoph Pohl\\[0.3ex]\normalsize}
\IEEEauthorblockA{MuSe - Munich IT Security Research Group\\
Munich University of Applied Sciences\\
Munich, Germany\\
Email: {\tt christoph.pohl0@hm.edu}}

\and
\IEEEauthorblockN{~\\[-0.4ex]\large Michael Meier\\[0.3ex]\normalsize}
\IEEEauthorblockA{Fraunhofer FKIE Cyber Defense\\
University of Bonn\\
Bonn, Germany\\
Email: {\tt michael.meier}\\{\tt @fkie.fraunhofer.de}}
\and
\IEEEauthorblockN{~\\[-0.4ex]\large Hans-Joachim Hof\\[0.3ex]\normalsize}
\IEEEauthorblockA{MuSe - Munich IT Security Research Group\\
Munich University of Applied Sciences\\
Munich, Germany\\
Email: {\tt hof@hm.edu}}}


%


\maketitle

\begin{abstract}
Honeypots are used in \itsec to detect and gather information about ongoing intrusions, e.g., by documenting the approach of an attacker. Honeypots do so by presenting an interactive system that seems just like a valid application to an attacker.
One of the main design goals of honeypots is to stay unnoticed by attackers as long as possible. The longer the intruder interacts with the honeypot, the more valuable information about the attack can be collected. Of course, another main goal of honeypots is to not open new vulnerabilities that attackers can exploit. Thus, it is necessary to harden the honeypot and the surrounding environment. This paper presents \apaten, a Linux Kernel Module (LKM) that is able to log, block and manipulate system calls based on preconfigurable conditions like Process ID (PID), User Id (UID), and many more. \apate can be used to build and harden High Interaction Honeypots. \apate can be configured using an integrated high level language. Thus, \apate is an important and easy to use building block for upcoming High Interaction Honeypots.

\end{abstract}


%
\IEEEpeerreviewmaketitle

\section{Introduction}
\label{introduction}

Honeypots are well known tools for \ids and \itsec research.
Usually, honeypots fall into one of two classes: Low Interaction Honeypots and High Interaction Honeypots. 
A Low Interaction Honeypot simulates attackable services, systems, or environments whereas a High Interaction Honeypot \cite{PohlHof:2013}\cite{PohlHof:2015} offers a real exploitable service, system, or environment. As in most cases a honeypot is not a productive system, every activity on a honeypot is either unintended use or an attack. 

When deploying a \hihn, it is necessary to harden the honeypot to avoid attackers gaining unintended control of the system running the honeypot. A \hih should by definition be exploitable, but it should prevent annoying or harmful operations on the honeypot system. Another important requirement for \hih is to log as much information as possible about the state of the system and about ongoing intrusions. Therefore, a \hih needs a highly flexible way to decide which information should be logged and which should not. \apate offers such a flexible way, using a high-level language for configuration. Also, it should be possible to log information on a as fine granular level as possible. Apate offers a logging on system call level. Manipulation of system calls, depending on user interaction or the system environment, is necessary to provide \hih functionalities. This allows the honeypot provider to present different environments depending on PID, UID (and many more), or system call parameters.
For example, the \hih provider is able to present one file structure to PID 42 and a completely different file structure to PID 43.
This manipulation can be used to decoy an attacker.  Furthermore, it can also be used to suppress harmful actions.
The honeypot admin is able to prevent execution of system call. Blocking a system call can be done by really blocking (not calling the real system call), or in a more sophisticated way. At last, it is necessary that \hih components (like the proposed LKM) should be hard to detect for intruders. This requirement calls for sophisticated technologies, already known from rootkits. For productive use, it is necessary that a \hih module has only low computational overhead. An attacker should not be 
able to detect a \hih by observing performance leaks.

Apate is a Linux Kernel module that fulfills all requirements mentioned above. Hence, it is an important building block for High Interaction Honeypots.

The rest of this paper is structured as follows: Section \ref{relwork} provides an overview on related work.
Section \ref{design} describes the design and implementation of Apate in detail. Section \ref{evaluation} shows the evaluation of Apate. Section \ref{conclusion} concludes the paper.

\section{Related Work}\label{relwork}
A well known honeypot tool, based on LKM for 2.6 Linux Kernel, is Sebek \cite{sebek2003}\cite{Balas:2003}.
Sebek is primarly used for logging purposes in \hihn. Thus, it provides several methods for detailed logging (like logging via network or GUI). In \cite{Holz:2005}\cite{Dornseif:2004uv}, ways to detect Sebek are described. Sebek does not provide the possibility to manipulate system calls, hence it does not offer such a fine-grain information logging as provided by Apate.

Another approach for monitoring systems is to use virtual machine introspection and system view reconstruction.
For example,  \cite{quebek:2010}, \cite{Lengyel:2012ve}, and \cite{Jiang:2007em} use this approach. Introspection realized on hardware level of the virtual machines offers a stealthier approach then Apate. However, Apate provides additional means to manipulate the behavior of system calls, which are not supported by \cite{quebek:2010}, \cite{Lengyel:2012ve}, and \cite{Jiang:2007em}, hence Apate is superior to these approaches.

SELinux \cite{smalley2001implementing} is a well known tool for inserting hooks at different locations inside the kernel.
Such an approach provides access control for critical kernel routines. SELinux can be controlled on a very fine granular level with an embedded configuration language. Although SELinux is very useful in hardening a kernel,  it is not designed for honeypot purposes. Especially, it lacks in the possibility to decoy the attacker using ``wrong'' information.

Grsecurity \cite{linkgrsec} with PAX \cite{linkpax} is similar to Apate.
However, it greatly differs in ease of deployment and ease of configuration \cite{fox2009selinux}. It also lacks in the possibility to decoy the attacker with ``wrong'' informations.

In conclusion, non of the mentioned related work fulfills all requirements listed in Section \ref{introduction}. Apate fulfills all requirements, hence is a useful building block for upcoming High Interaction Honeypots.

\section{Design and implementation}\label{design}

\begin{figure}[!htp]
\centering
\begin{tikzpicture}
\node (syscall) [process, minimum width=4cm] {Call Syscall};
\node (hook) [process, below of=syscall, minimum width=4cm] {Interception (Hook)};
\node (orig) [process, below of=hook, minimum width=4cm] {Original Syscall};
\node (hook2) [process, below of=orig, minimum width=4cm] {Interception (Hook)};
\node (rules) [process, right=.35cm of orig] {Rules};
\node (return) [process, below of=hook2, minimum width=4cm] {Return};

\draw [->](syscall) -- (hook);  
\draw [->](hook) -- (orig); 
\draw [->](orig) -- (hook2);
\draw [->](hook2) -- (return);
\draw [<->](hook.east)--(rules.west);
\draw [<->](hook2.east)--(rules.west);
\end{tikzpicture} 
\caption{interception strategy of Apate} 
\label{fig:concept}
\end{figure}
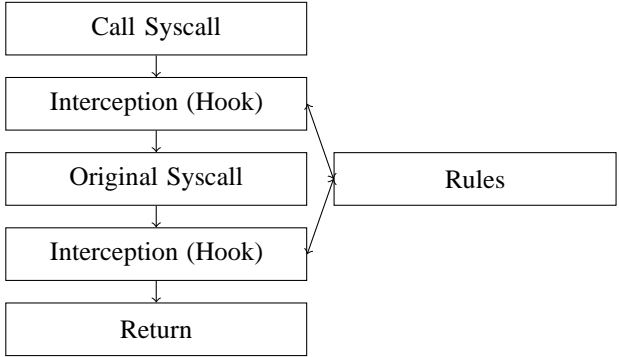

Apate intercepts system calls and allows to execute custom code in these calls. Figure \ref{fig:concept} shows the interception strategy of Apate.

Apate does not manipulate the syscall table to prevent detection (see Subsection \ref{stealth} for details). Apate intercepts the syscall within the syscall target, i.e., the real syscall address is called but Apate jumps immediately to the interception routine (after consuming some decoy assembler code). The hook decides on the action to invoke, based on the rules for this system call. Within this action, it is able to manipulate, block and/or log a system call.
The following hooks are implemented in \apate:
\begin{itemize}
\item sys\_open, sys\_close, \syscall{open}
\item sys\_read, sys\_write, sys\_unlink
\item sys\_execve
\item sys\_getpid, sys\_getuid
\item sys\_mkdir, sys\_rmdir
\item sys\_getdents
\end{itemize} 

This paper focuses on the usage of system calls that are related to File IO and execution control as these system calls are usefull for hardening High Interaction Honeypots.  

\subsection{Configuration}\label{configuration}
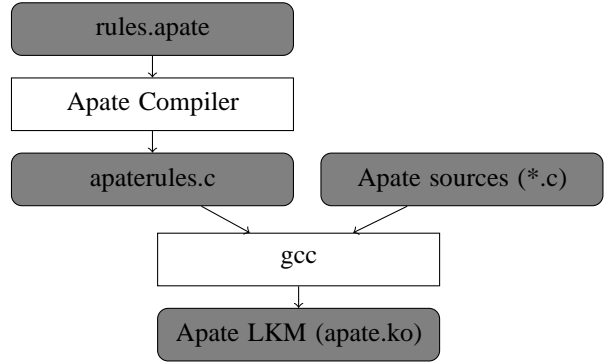
\begin{figure}[!htp]
\centering
\begin{tikzpicture}

\node (rules) [data] {rules.apate};
\node (acompiler) [process, below of=rules] {Apate Compiler};
\node (rulesc) [data, below of=acompiler] {apaterules.c};
\node (osource) [data, right=1em of rulesc] {Apate sources (*.c)};

\node (gcc) [process, below right=1em and -1.82cm of rulesc] {gcc};
\node (lkm) [data, below of=gcc] {Apate LKM (apate.ko)};

\draw [->](rules) -- (acompiler); 
\draw [->](acompiler) -- (rulesc); 
\draw [->](rulesc) -- (gcc); 
\draw [->](osource) -- (gcc); 
\draw [->](gcc)--(lkm);

\end{tikzpicture} 
\caption{Configuration workflow of Apate} 
\label{fig:config}
\end{figure}

Apate can be configured in a very flexible way as can be seen in Figure \ref{fig:config}.
The configuration file \itan{rules.apate}, written in a high level language (see section\ref{subsec:conf} for details), gets compiled by the \apate compiler, resulting in the file
\itan{apaterules.c}. Together with the original source code, the compiler generates the \apate LKM.
The resulting LKM can be loaded into kernel with common insmod util.
Once loaded, the ruleset is active. 

The configuration consists of rulesets.
A ruleset is an ordered list of rules.
A system call gets intercepted when one or more rules match.
One system call can have more than one matching rule with different decision parameters.
There are three major types of decision parameters:
\begin{itemize}
\item Parameters that are system call independent like PID, UID, SSID (in fact every variable from \ita{struct task\_struct}\cite{linkerntask} could be used for conditions).
\item Parameters that are dependent to the specified system call. Often, these are function parameters like paths.
\item Parameters that are defined by functions. This decision parameter allows to build reactive systems. For example, one can define a condition which reads some file, whenever the file contains a keyword the condition could be true.
\end{itemize}

Rules are defined as stated below:

Let be $true=1$ and $false=0$. Let $c(a,b)$ be a condition such that:
\begin{formelnon}
&C:A \times B \rightarrow \{0,1\} \\
&(a,b)\mapsto c(a,b)
\end{formelnon}
, where $a\in A$ and $b\in B$ are two parameters, which are used by $c()$ for the calculation of the condition. The parameters are further called decision parameters. For example, a decision parameter could be the path of the system call \itan{\syscalln{open}()} and the related condition is  \itan{if (param[0] == ``/etc/passwd'') ? 1 : 0} where $a=param[0]$ and $b=``/etc/passwd''$ . This rule matches system calls trying to get access to the file $``/etc/passwd''$

Let $cb(d,e,f)$ be a condition block.
A condition block calculates the result of conditions or other conditionblocks with \itan{AND} or \itan{OR}.
A condition block uses two parameters $d$ and $e$ and an operator $f$.
$d$ and $e$ can be the result of any $c(a,b)$ or another $cb(d,e,f)$. 

$CB$ is the set of all possible condition blocks: 
\begin{formelnon}
&CB:\{0,1\} \times \{0,1\} \times \{AND, OR\} \rightarrow \{0,1\} \\
&(d,e,f) \mapsto cb(d,e,f)
\end{formelnon}

Further, the set of all conditions and condition blocks is $AC$ such that
\begin{formelnon}
AC = C \cup CB
\end{formelnon}

For $cb(d,e,f)$ let $d,e \in AC $ and $f \in \{2,4\}$.
When $f=2$ the operator $AND$ will be used.
When $f=4$ the operator $OR$ will be used.
$c^*(a,b)$ is a condition that returns always true. 
The second parameter in the conditionblock can be neutralised with $cb(d,c^*(0,0),2)$

This leads to the definition
\begin{formelnon}
cb(d,e,f) = 
\begin{cases}
1 & \text{if } (d+e) * f \geq 4\\
0 & \text{other}
\end{cases}
\end{formelnon}

This definition makes it possible to group different conditions and to be aware of precedences.

Let $A$ be the set of atomic actions. An atomic action is a function that provides only one single functionality.
For example, an atomic action can be the redirection of a system call. An action $a \in A$ falls in one of three groups:
\begin{itemize}
\item Manipulating actions
\item Logging actions
\item Blocking or emergency exit actions
\end{itemize}
Let $AS$ be an orderd list of actions. The index function $i(x)$ assigns an index to each element $x \in AS$, hence

\begin{formelnon}
AS = \{x \in AS| 0 < i(x-1) < i(x)\} 
\end{formelnon}

Let $AAS$ be the lists of all actions. A rule $r^{g,h}$ consists of one condition block $g \in CB$ and an action set $h \in ASS$. 
Let $R$ be the set of all rules.
Whenever the condition block returns $1$, the action set $h$ is started. 

Let $RS$ be a list of sorted rules ($RS \in R$). Each element of $RS$ has a flag $fl$. A flag is defined as 
\begin{formelnon}
fl \in \{exit=1,\neg exit=0\}
\end{formelnon}
When a system call gets called, all rules in $RS$ are calculated beginning with the first rule in $RS$ and until a rule is in state $true$ and $fl=1$. 


Using the definitions above, a highly configurable system could be build. Including some basic predefined conditions enhances convenience, e.g., equality checks for integer, floats or strings.

\subsection{Configuration High Level Language}\label{subsec:conf}
The configuration language implements two main requirements:
first, the configuration language should be flexible, including the ability to reuse patterns, store variables, calculate with operators, embed external functions, define functions, and use decision statements.  This allows to use the language to describe even very complex scenarios. Second, the configuration language should provide a transparent way to define rules, related to honeypots (or in scope of this paper to control and manipulate system calls). To deal with these requirements, the \apate language combines concepts known from functional programming (in this case Haskell \cite{linkhaskell}) with a concept well known from packet filter configuration (in this case pf \cite{linkpf}).

This Section gives a brief introduction to the important parts of the language. For the sake of  clarity, some convenience features of the Apate language ( e.g., embedded C, self defined functions, loops) are omitted. 

Listing \ref{lst:ex} shows some example source code for the \apate language.
\begin{figure}[!htp]
\begin{lstlisting}
define c1,c2,c3 as condition
define r1,r2 as rule
define a1,a2 as action
define cb1 as conditionblock
define rc1 as rulechain
define sy1 as syscall

let c1 be testforpname
let c2 be testforparam
let c3 be testforuid
let a1 be manipulateparam
let a2 be log
let sy1 be sys_open

let cb1 be {(c1("mysql") && c2(0;"/var/\
   lib/mysql/*"))}

let r1 be {cb1->a1(0;"/var/lib/mysql/*" \
  ;"/honey/mysql/")}
let r2 be {{c3(">",0)}->a2()}
let rc1 be {r2,:r1} // :defines exit

bind rc1 to sy1
\end{lstlisting}
\caption{Example Sourcecode \apate language}
\label{lst:ex}
\end{figure}

The first block with the \verb|define| statements binds variables to different types (like condition, rules, or functions).
The code block with the \verb|let| statements points these variables to values or functions.
In this case, it defines 3 conditions (\verb|c1,c2,c3|).
\verb|c1| will test the actual process name against another string.
\verb|c2| tests if a param of the actual syscall is equal to a given value.
\verb|c3| tests if the actual uid is equal to a given value.
\verb|a1,a2| are actions. \verb|a1| manipulates a parameter of the actual system call. 
\verb|a2| logs a system call.
The variable \verb|cb1| represents a condition block.
Its \verb|let| assignment also shows that it is possible to write nested variable assignements.
In this case, the conditions \verb|c1,c2| are combined with \verb|&&| (\itan{AND}).
In the same line, the conditions \verb|c1,c2| gets assigned with parameters.
In this case the condition \verb|c1| checks if the current parent process is the mysql-Process.
\verb|c2| checks if the first parameter (\verb|0|) of the current system call is equal to \ita{/var/lib/mysql/*}.
The asterisk describes a wildcard function.
The rule assignment for \verb|let r1 be...| binds a conditionblock to an action.
In this case, it means whenever the conditionblock returns true the action \verb|a1| rewrites the first param of the current system call.
It replaces \ita{/var/lib/mysql} with \ita{/honey/mysql}.
The rule \verb|r2| logs the current system call whenever the current UID is greater than 0.
A ruleset (rulechain) \verb|rc1| is assigned with \verb|r2,r1|.
The \verb|r1| rule is also assigned as exit rule (\verb|..:r1..|).
When this rule fires, no further rules will be called. In the last line, the rule chain \verb|rc1| is bound to the system call \itan{sys\_open}.

In conclusion, when the system call \syscall{open} gets called, the parent process is the mysql process and the system call parameter (in this case the path which should be opened) begins with \itan{/var/lib/mysql/*}, this syscall gets manipulated and the syscall will open a file under \ita{/honey/mysql/...}.
The second rule means that every call for \syscall{open} will be logged, except when the root user calls this system call.

\subsection{Manipulation of System Calls}\label{manipulation}
If a rule matches, the corresponding action chain gets called to manipulate the original system call. An action chain has a length $l$ with $1 \leq l < n$.

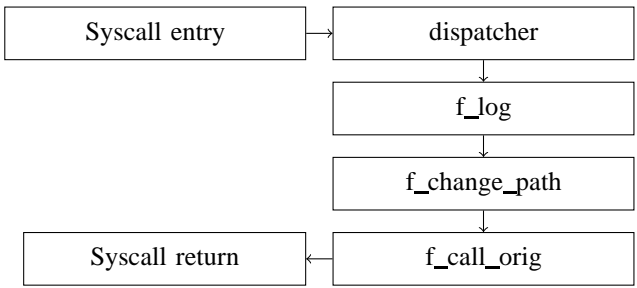
\begin{figure}[!htp]
\centering
\begin{tikzpicture}
\node (syscall) [process, minimum width=4cm] {Syscall entry};
\node (rule) [process, right=.35cm of syscall,minimum width=4cm] {dispatcher};
\node (flog) [process, below of=rule,minimum width=4cm] {f\_log};
\node (fman) [process, below of=flog, minimum width=4cm] {f\_change\_path};
\node (orig) [process, below of=fman, minimum width=4cm] {f\_call\_orig};
\node (syscallex) [process, left=.35 of orig] {Syscall return};

\draw [->](syscall) -- (rule); 
\draw [->](rule) -- (flog); 
\draw [->](flog) -- (fman); 
\draw [->](fman)--(orig);
\draw [->](orig)--(syscallex);
\end{tikzpicture} 
\caption{Conceptual Manipulation Strategy} 
\label{fig:manipulation}
\end{figure}

Figure \ref{fig:manipulation} shows an example for the manipulation strategy. Functions prefixed with \ita{f\_} are actions.

The dispatcher represents the rule engine, deciding which action chain should be used.
In this example the first action logs the system call.
The next action manipulates some parameter like a path or anything else.
The \ita{f\_call\_orig} calls the original system call with the manipulated parameter.
The result gets returned to the callee.

Technically, one action is a function that consumes all system call parameters, including the current \itan{struct task\_struct} and a pointer to the \ita{syscall\_result} variable.
Each function returns an Integer, indicating wether the function call has been successful or not.
Whenever a function returns an error, the action chain gets disrupted and an error routine is called.
Finally, the hook returns the \itan{syscall\_result}.
In case of an error the system call returns a system call dependent error.
 
\subsection{Hiding Hooks and LKM}\label{stealth}
An attacker should not be able to detect \apaten, otherwise Apate would not be suitable for High Interaction Honeypots. As hiding software in all use cases is very difficult, Apate must at least hide itself until the effort of detection of Apate is unreasonable high for an intruder. This requires to define which effort is unreasonable high for an attacker and which is not.
The following actions are defined as reasonable for an attacker, hence should be prevented:
\begin{itemize}
\item Testing for module presence with standard utils like \ita{lsmod,modinfo,...} or misleading errors when using \ita{insmod} and similar tools
\item Testing for presence of module in \ita{/proc/module} and \ita{/sys/module}
\item Testing for presence of \apate related logfiles, configurations, and other artifacts
\end{itemize}

To hide Apate, it is necessary to remove the module from the module list.
Simplified, all modules are represented in a global linked list.
By using \begin{verbatim}list_del_init(&__this_module.list);\end{verbatim} the module is removed and therefore invisible.
To hide from the \ita{/sys/module} \apate uses \begin{verbatim}kobject_del(&THIS_MODULE->mkobj.kobj);\end{verbatim} to remove itself from this representation.
With these modification, the module is invisible to standard utils (they use \ita{/proc/module}) and in \ita{/sys/module}.
These technologies are also well known rootkit technologies see for example \cite{ligh2014art}\cite{linkrootkit}.

\apate does not use any configuration files beyond the configured rules. The high level language should be deleted by the honeypot admin after its compilation into \apaten. Hence, Apate cannot be identified by an attacker looking for configuration files.

Apate is used to cloak logfiles: predefined rules in Apate prevent all users to see, read, or write Apate logfiles.
To gain access to the logfile, a system administrator need to restart the host system without the honeypot. 

To detect a hook, an intruder needs to analyze physical memory. \apate makes it hard to load a new module into the kernel.
It prevents to load another kernel module by overriding the flag that controls the module loading ability.
Beyond the possibilities of \apaten, the honeypot admin can harden the host system to ensure that this dumping has a high effort for an intruder.

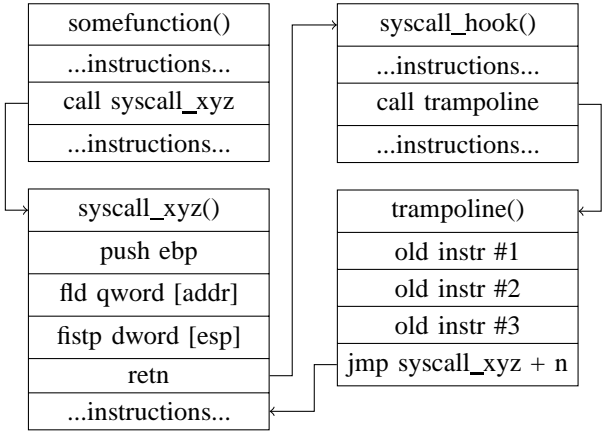
\begin{figure}[!htp]
\centering
\begin{tikzpicture}[stack/.style={rectangle split, rectangle split parts=#1,draw, anchor=center, minimum width=3.2cm}]
\node(calling)[stack=4]  {
\nodepart{one}   somefunction()     
\nodepart{two}   ...instructions...     
\nodepart{three} call syscall\_xyz      
\nodepart{four}  ...instructions... 
};

\node(syscall)[stack=6, below=1em] at (calling.four south)  {
\nodepart{one}   syscall\_xyz()     
\nodepart{two}   push ebp     
\nodepart{three} fld qword [addr]      
\nodepart{four}  fistp dword [esp]
\nodepart{five}  retn
\nodepart{six}   ...instructions...
};

\node(hook)[stack=4, below right=0cm and 2.5cm of calling.north]   {
\nodepart{one}   syscall\_hook()     
\nodepart{two}   ...instructions...     
\nodepart{three} call trampoline
\nodepart{four}  ...instructions...
};

\node(trampoline)[stack=5, below=1em of hook]  {
\nodepart{one}   trampoline()     
\nodepart{two}   old instr \#1     
\nodepart{three} old instr \#2      
\nodepart{four}  old instr \#3
\nodepart{five}  jmp syscall\_xyz + n
};


\draw [->, bend angle=45, bend left] (calling.three west) --++ (-.3,0) |- (syscall.one west);
\draw [->, bend angle=45, bend left] (trampoline.five west) --++ (-.3,0) |- (syscall.six east);
\draw [->] (syscall.five east) --++(.3,0) |- (hook.one west);
\draw [->] (hook.three east) --++ (.3,0) |- (trampoline.one east);
\end{tikzpicture} 
\caption{Hooking using a so-called trampoline} 
\label{fig:trampoline}
\end{figure}

\apate has different opportunities to insert hooks into system calls. By default, \apate changes the function pointer in the system call table. This is sufficient as long as the intruder has no possibility to compare the original table with the hooked table. If this is not enough protection, the admin can decide to harden the system with some anti-rootkit technologies. This makes it impossible for \apate to overwrite the jump points. Figure \ref{fig:trampoline} shows the alternative hooking technology.
This technology is well known from Windows and Linux rootkits.
During the hooking process, \apate stores the first $n$ bytes of the target system call function.
The stored commands will be copied to a trampoline function.
Instead of the original commands, \apate injects a jump operation.
This lets the process jump into the hooking function immediately after entering the original system call function.
Whenever the hooking function calls the original system call, it calls the trampoline function. 
The original code is executed, then the trampoline function lets the process jump into the original function with an offset of $n$ bytes. The trampoline is a feature to obfuscate the hook for rootkit detection tools and uses live patching technologies.
Thus, it can be detected with core dump disassembling.
This is out of scope of this paper as it is assumed that the effort to detect the honeypot with disassembling tools is too high.  

\section{Evaluation}\label{evaluation}
There are three major goals for \apaten.
The first goal is to provide a highly flexible configuration.
Although the proposed configuration system works well and is suitable for \hihn s, it must be ensured that every possible combination of rulesets and actions can be described.
This means that the configuration language must be turing complete.
For this, it is determined that an action $a$ is able to decide which rule from the ruleset should be invoked next.
This means it can jump to any other rule from a given ruleset.
It is also determined that \apate has an array (in this case an impossible array with infinite indices which can hold any other type (like actions, rules, other defined variables or anything else)).
Technically, \apate has a register and a stack.
Last, it is determined that an action is able to fill or read any index of this array.
Together with actions for calculations and conditions for jump decisions, the system is turing complete.
In fact any action is just a C-Function and the \ita{define} statement creates variables. 

The second goal is to provide a system which achieves a suitable level of stealthiness.  As described in Subsection \ref{stealth}, the system hides itself from common util tools like \ita{lsmod, modinfo, modprobe, insmod}.
\apate is also not available in \ita{/proc/module} or \ita{/sys/module}. To test for presence in any logfile a simple grep command with typical signatures for \apate (Simplified each log entry or configuration includes the string ``apate'', thus it is easy to detect it) is fired on the full system.
However with proper rules these log entries are not visible by standard system commands.

The third goal is that Apate should be efficient. Performance tests should assure that \apate is able to serve under productive usage scenarios. The most important performance factor is the overhead of logging. To evaluate the performance of \apate in a productive scenario, the execution time of \syscall{open}, \syscall{write},\syscall{read}, and \syscall{close} are measured.
The \syscall{open} and \syscall{close} get called just once a file is opened or closed.
The \syscall{write} and \syscall{read} get called more often (under the condition that heavy writing will be done on the system).
Thus, the test pattern concentrates on \syscall{write} and \syscall{read}. For the performance evaluation, data is copied from one file to another using increasing file lengths. This will be done for 100 times for each file size.
The source file is generated on the fly from \ita{/dev/random} before each copy command.
After each successful copy command the target file is deleted.
A Gentoo 64 Bit system with 32 GB Ram and 16 Cores is used for all performance tests. The kernel is optimized by disabling unnecessary drivers and by enabling some debugging flags. One source file is generated for each size with random bits and a length of $l(file)$ bytes. Let the size of the file be:
\begin{formelnon}
 0 < l < 1,000,000,000
\end{formelnon}
and 
\begin{formelnon}
l_n(file) =  
\begin{cases}
l_{n-1} + 1 & \text{if } l_n < 1,000,000\\
l_{n-1} + 1,000 & \text{if } 1,000,000 \leq l_n\\
& \wedge l_n < 100,000,000\\
l_{n-2} + 1,000,000 & \text{if } 100,000,000 \leq l_n \\
& \wedge l_n \leq 1,000,000,000
\end{cases}
\end{formelnon}

Four different settings were tested. 

The first setting ($m_1$) is used as reference setting. It does not use any interception. 

The second setting, $m_2$, uses only one rule which always returns true. 
The related action set calls the origin system call and logs this action.
This is the shortest way in \apate to provide logging functionality.
This testing is used to evaluate the logging overhead of Apate.
 
The third and fourth setting, $m_3$ and $m_4$, evaluate the influence of rules.
Each rule consists of 50 conditions with $\{c_0,c_1,\dots,c_{50}\}$ where each condition is combined with an \ita{and} statement.
The last condition returns false.
Each test uses 50 rules.
The last condition in rule number 50 (last rule) returns true.
Overall, each system call passes 2500 conditions.
This triggers an action set that will call the original system call ($m_3$ and $m_4$) and then logs this action (only $m_4$).


Table \ref{tab:perf} shows the results of the performance evaluation. The \ita{sd}-row shows the standard deviation, \ita{var} shows the variance, and \ita{Iqr} shows the interquartile range.

\captionsetup{font={footnotesize,sc},justification=centering,labelsep=period}%
\begin{table}[!t]
\renewcommand{\arraystretch}{1.3}
\caption{Performance Measurement}
\label{tab:perf}
\centering
\begin{tabular}{l | l l l l}
\hline
\bfseries Measurement & \bfseries $m_1$ & \bfseries $m_2$ & \bfseries $m_3$ & \bfseries $m_4$\\
\hline\hline
Measurements & 110,800 & 110,800 & 110,800 & 110,800\\
Unique Filesizes & 1,108 & 1,108 & 1,108 & 1,108\\
sd(runtime sec) & 0.1066& 0.2421& & 0.2452\\
var(runtime sec) & 0.0114& 0.0586& & 0.0601\\
iqr(runtime sec) & 0.0010& 0.0026& & 0.0023\\
\hline
\end{tabular}
\end{table}
\captionsetup{font={footnotesize,rm},justification=centering,labelsep=period}%


\begin{figure}
 \centering
\input{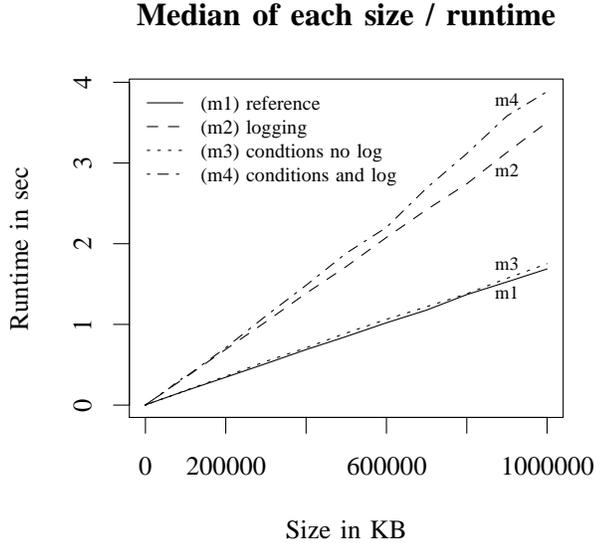}  
  \caption{Relation between runtime/filesize/rules \syscall{open}}\label{fig:perf}
\end{figure}

Figure \ref{fig:perf} shows the correlation between file size and runtime. For every curve the median of the measured runtimes for each unique file size is connected with a line. The $m_1$ curve shows the reference setting. The $m_2$ curve shows that the logging component has a big influence on performance. Each syscall and its values were logged. Each log was sent to another server using UDP. Gentoo uses a buffer with 65,365 Byte. To copy a file with one Gigabyte it needs 32,720 syscalls. This explains the overhead of $m_2$ and $m_4$. The $m_3$ curve shows that the rule engine works with just a small overhead when only conditions get processed. For one measurement with a file size of one Gigabyte, the engine processed 81,800,000 conditions.
However, to copy a file with less than 65,365 Byte only 4 syscalls are passed and therefore only 10,000 conditions gets processed.

In conclusion, these measurements shows that it is possible to build a syscall interception framework which is able to provide proper configuration with acceptable overhead. The evaluation does not show a single case that prevents a productive usage of Apate. 

\section{Conclusion}\label{conclusion}
This paper presented Apate, a Linux Kernel Module for hardening High Interaction Honeypots. Apate works on a system call level, is able to log, block and manipulate these calls, and uses an easy to use yet powerful configuration language. The evaluation shows that Apate has a moderate performance overhead and can be used in productive honeypot systems. \apate is also stealthy enough for most common usage scenarios. Overall, Apate is an ideal basis and important building block for upcoming High Interaction Honeypot Systems.



%
%
%

\bibliographystyle{IEEEtran}
\bibliography{bib}

\end{document}